\documentclass[12pt,aps]{article}
\usepackage{graphicx}
\usepackage{amsmath}
\usepackage{mathrsfs}

\newcommand{ \RN}{Reissner-N\"{o}rdstrom }
\newcommand{\CH}{Cauchy Horizon }
\newcommand{\EH}{Event Horizon }

\newcommand{\half}{\frac{1}{2}}
\newcommand{\EQ}[1]{Eq.(\ref{EQ #1})}  % Equation referencing
\newcommand{\FIG}[1]{Fig.(\ref{FIG #1})}  % figure referencing
\newcommand{\verylongrightarrow}{-\!\!\!-\!\!\!-\!\!\!-\!\!\!-\!\!\!\!\!\longrightarrow}

\begin{document}
\title{On the Collapse of Charged Scalar Fields}
\author{Yonatan Oren and Tsvi Piran\\
\small
  \textit{
    The Racah Institute of Physics, The Hebrew University,
    Jerusalem, Israel, 91904}}
\date{}
\maketitle
\begin{abstract}
We explore numerically the evolution of a collapsing spherical
shell of charged, massless scalar field. We obtain an external \RN
space-time, and an inner space-time that is bounded by a
singularity on the Cauchy Horizon. We compare these results with
previous analysis and discuss some of the numerical problems
encountered.
\end{abstract}

\section{Introduction}
The most general analytic, stationary,  and asymptotically flat
solution of the Einstein equations is the Kerr-Newman metric,
which is characterized by three parameters - mass, charge and
angular momentum. The no-hair theorem \cite{Wheeler} suggests that
at asymptotically late times after the collapse of a massive body
to a black hole, the external metric of the black hole relaxes to
the Kerr-Newman metric. However, the final state of the inner
region is not generally determined and has been the subject of
research for some time. Also, it is of interest, especially to
gravitational wave astronomers, to predict the specific details of
the relaxation of the ``hair''.

This appearance of a time-like singularity and a ``chain'' of
asymptotically flat regions in the maximally extended \RN metric
(\FIG{penrose RN}) (and also in the Kerr metric) opened up the
discussion of the existence and stability of ``wormholes''.
The \RN metric is problematic as a realistic black hole
picture because the part of space-time that lies in the future of
the \CH is unpredictable by a set of Cauchy initial conditions
given on a space-like surface in the asymptotically flat region.
This happens because events in the region beyond the \CH are in
causal connection to the time-like singularity.

\begin{figure}[t]
\begin{center}
\includegraphics[width=7cm,height=9cm]{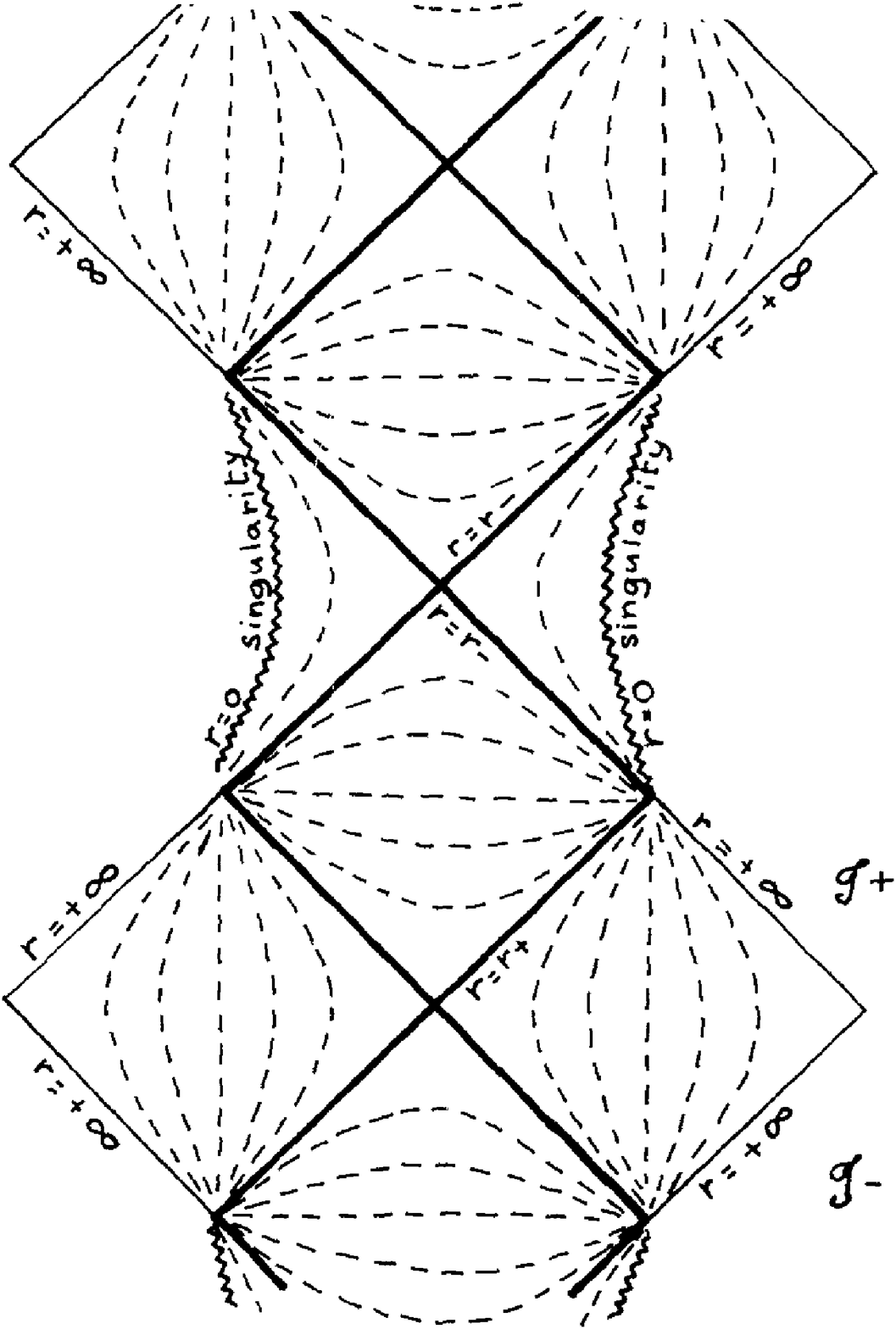}
\end{center}
\caption{A Penrose diagram for the maximally extended \RN
space-time. The \CH is marked as $r=r_-$, The \EH as $r=r_+$.
Taken from \cite{LesHouches}. } \label{FIG penrose RN}

\end{figure}

Consequently we would expect that a in a dynamic collapse
the space-time will not extend beyond the Cauchy Horizon, but
would rather be bounded there by a singularity of some sort. This
scenario was predicted by Penrose \cite{Penrose_blueshift} as
early as 1968. The physical mechanism leading to this singularity
is known as the ``Mass Inflation'' scenario \cite{Poisson}, and is
connected to the infinite blueshift of in-falling radiation on the
Cauchy Horizon.

Previous results \cite{Hiscock,Ori1,Ori2} concerning the stability
of the cauchy horizon were acquired either by linear perturbation
theory or by numerically evolving a neutral perturbation
collapsing on a preexisting \RN black hole
\cite{Hamade,Brady,Burko}. The latter model is relatively easy to
solve numerically since it does not contain charges and dynamical
electromagnetic fields, but on the other hand it does not give the
complete picture of the evolution of the black hole and the \CH
beginning with a flat space time. Previous work dealing with a
fully dynamical charged collapse model
\cite{HPtail1,HPtail2,HPtail3,HPmassinf} focused mainly on the
behavior near the origin and outside the horizon, i.e at
relatively large values of retarded time but moderate values of
advanced time, on the order $10M$. These works indicated that in a
charged collapse the \CH is replace by a weak, null singular
segment connected to a space-like singularity at the origin
(\FIG{RN collapse}). In this work we follow the evolution
of a charged matter shell that collapses to form a black hole, and
determine the final structure of the inner metric, and the \CH
at \textit{large} values of advanced time, deep inside the Black
Hole interior.
\begin{figure}[h]
\begin{center}
\includegraphics[width=4cm,height=6cm]{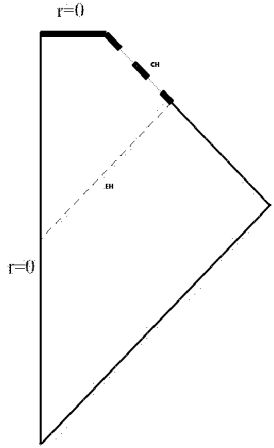}
\end{center}
\caption{A Penrose diagram for a charged collapse scenario. The dashed lines
indicate singularities along the origin and Cauchy Horizon.Taken from \cite{LesHouches}. \label{FIG RN
collapse}}
\end{figure}

In section 2, we formulate our basic equations, starting from the
gauge invariant Lagrangian of the complex scalar field and casting
these equations into the double-null coordinate system. In section
3 we discuss the numerical scheme, it's accuracy and the
difficulties that arise in this kind of calculation. In section 4
we discuss the numerical results and their implications regarding
several aspects of black hole physics, such as Mass Inflation,
Cosmic Censorship, Wave tails and the No-Hair theorem, and the
nature of the \CH singularity.

\section{Formulation of the problem}

\subsection{The equations}
We begin with the gauge-invariant Lagrangian for the complex
scalar field $\psi$ and the electromagnetic gauge field $A_\mu$
\cite{Hawking}:
\begin{equation}\label{EQ lagrangian}
\mathscr{L}=-(\psi_{;a}+ieA_a\psi)g^{ab}(\bar\psi_{;b}-ieA_b\bar\psi)-\frac{m^2}{\hbar^2}\bar\psi
\psi-\frac{1}{8\pi}F_{ab}F_{cd}g^{ac}g^{bd},
\end{equation}
where we define $F_{\mu\nu}=A_{\nu;\mu}-A_{\mu;\nu}$. We are dealing with a
massless field so we will omit the middle term from now on. From this
lagrangian we can derive the Euler-Lagrange equations for the scalar and
electromagnetic field, and also the Energy-Momentum tensor to be used in the
Einstein equations. The wave equations for the scalar field are:
\begin{equation}\label{EQ wave}
\psi_{;ab}g^{ab}+ieA_a g^{ab}(2\psi_{;b}+ieA_b\psi)+ieA_{a;b}
    g^{ab}\psi=0,
\end{equation}
and its complex conjugate. The Maxwell equations take the form:
\begin{equation}\label{EQ maxwell}
\frac{1}{2\pi} F_{ab;c} g^{bc}+ie\psi(\bar\psi_{;a}-ieA_a\bar\psi)-
ie\bar\psi(\psi_{;a}+ieA_a\psi) = 0.
\end{equation}
The Energy-Momentum tensor given by:
\begin{eqnarray}\label{EQ energytensor}
T_{ab} &=&
\frac{1}{2}(\psi{;a}\bar\psi_{;b}+\bar\psi_{;a}\psi_{;b})+
\frac{1}{2}(-\psi_{;a}ieA_b\bar\psi+\bar\psi_{;b}ieA_a\psi+\bar\psi_{;a}ieA_b\psi-\psi_{;b}ieA_a\bar\psi)+\nonumber\\
&&\frac{1}{4\pi}F_{ac}F_{bd}g^{cd}+e^2A_aA_b\psi\bar\psi+
\frac{1}{2}\mathscr{L}g_{ab}.
\end{eqnarray}

We choose the (2+2) spherically symmetric, double null coordinate
system. The line element in these coordinates can be written as:
\begin{equation}\label{EQ metric}
ds^2 = -\alpha(u,v)^2 dudv + r(u,v)^2 d\Omega^2,
\end{equation}
with r being the area coordinate.
Note that any coordinate gauge of the form $u\rightarrow f(u), v\rightarrow
g(v)$ will preserve the null character of u and v. We will fix this gauge
freedom later when discussing the initial conditions.

We need also to fix the gauge freedom of the electromagnetic field
$A_\mu$. In our symmetry and coordinate choice the only
non-vanishing components of F are $F_{uv}=-F_{vu}$. This means
that only $A_u$ and $A_v$ may be non-zero. We can eliminate one of
these using the electromagnetic field gauge freedom, $A_\mu
\rightarrow A_\mu + \Phi_{;\mu}$, where $\Phi$ is an arbitrary
scalar function. We choose to eliminate $A_v$ by taking
$\Phi=-\int A_v dv$, and so we are left with $A_\mu=(A_u,0,0,0)$.
We shall call this quantity $a\equiv A_u$ from now on.

We now proceed to write our equations in an explicit form. We
begin with the Maxwell equations. Since we reduced the potential
to one component using the gauge freedom, we need only one
equation for this field. \EQ{maxwell} is a vector equation with
two non-vanishing components u and v, and we may choose either one
to evolve the electromagnetic field. We choose the v component,
and will use the u component later when checking charge
conservation. The v component then becomes (we are using from now
on the convention $Z_\mu=\frac{\partial Z}{\partial x^\mu}$):
\begin{equation}\label{EQ maxwell v}
(\frac{r^2 a_v }{\alpha^2})_v+\half ier^2\pi (\bar\psi \psi_v-\psi
\bar\psi_v)=0.
\end{equation}
We define:
\begin{equation}\label{EQ q definition}
q\equiv \frac{2 r^2 a_v}{\alpha^2},
\end{equation}
and thus separate the $2^{nd}$ order equation for {\it a} into two simpler
$1^{st}$ order equations:
\begin{eqnarray}\label{EQ maxwell A}
a_v &=& \frac{\alpha^2 q}{2 r^2} \nonumber \\
q_v &=& 2\pi iear^2 (\psi \bar\psi_v-\bar\psi \psi_v).
\end{eqnarray}
The function $q(u,v)$ defined above is the amount of charge within
(i.e. at smaller radii than) a sphere of radius $r(u,v)$ on some
space-like hypersurface that contains $(u,v)$.

The mutually independent elements of $G_{\mu \nu}$ read:
\begin{eqnarray}\label{EQ Gtensor component}
    G_{uu} &=& 2\frac{2\alpha_u r_u-\alpha r_{uu}}{\alpha r} \nonumber\\
    G_{vv} &=& 2\frac{2\alpha_v r_v-\alpha r_{vv}}{\alpha r} \nonumber\\
    G_{uv} &=& \frac{\alpha^2+4r_ur_v+4rr_{uv}}{2r^2} \nonumber\\
    G_{\theta \theta} &=& 4\frac{r^2\alpha_u\alpha_v-r^2\alpha
    \alpha_{uv}-\alpha^2 r r_{uv}}{\alpha^4},
\end{eqnarray}
Combining \EQ{Gtensor component} and the Einstein tensor $G_{\mu
\nu}$ we arrive at the field equations:
\begin{eqnarray}\label{EQ field A}
&&\hspace{-20mm}r_{vv}-2r_v\frac{\alpha_v}{\alpha}+4\pi r \bar\psi_v \psi_v=0 \nonumber \nonumber \\
&&\hspace{-20mm}r_{uu}-2r_u\frac{\alpha_u}{\alpha}+4\pi r (\bar\psi_u
\psi_u+iea(\psi
\bar\psi_u-\bar\psi \psi_u)+e^2 a^2 \bar\psi \psi)=0 \nonumber \nonumber \\
&&\hspace{-20mm}rr_{uv}+r_u r_v+\frac{\alpha^2}{4}-\frac{\alpha^2 q^2}{4r^2}=0 \nonumber \\
&&\hspace{-20mm}\frac{\alpha_{uv}}{\alpha}-\frac{\alpha_u
\alpha_v}{\alpha^2}+\frac{r_{uv}}{r} +\frac{\alpha^2 q^2}{4
r^4}+2\pi(\bar\psi_u \psi_v+\bar\psi_v \psi_u) +2\pi iea(\psi
\bar\psi_v-\bar\psi \psi_v)=0
\end{eqnarray}
Finally we evaluate the wave equation and arrive at the form:
\begin{equation}\label{EQ wave A}
r\psi_{uv}+r_u\psi_v+r_v\psi_u+iear\psi_v+iear_v\psi+ie\frac{\alpha^2
q}{4r}\psi=0.
\end{equation}

We are now in a position to write down the full set of equations
to be solved. For this purpose we introduce some new notations:
\begin{eqnarray}\label{EQ notations}
&&s\equiv \sqrt{4\pi}\psi \nonumber \\
&&w\equiv s_u \ ;\ z\equiv s_v \nonumber \\
&&f\equiv r_u \ ;\ g\equiv r_v
\end{eqnarray}
Rearranging and substituting we arrive at the following set of equations:
\begin{eqnarray}\label{EQ equations B}
E1:&& r r_{uv}+fg+\frac{\alpha^2}{4}-\frac{\alpha^2 q^2}{4r^2}=0 \nonumber \\
E2:&&
(ln\alpha)_{uv}-\frac{fg}{r^2}-\frac{\alpha^2}{4r^2}+\frac{\alpha^2q^2}{2r^4}
+\frac{1}{2}(w\bar{z}+\bar{w}z)+\frac{1}{2}iea(s\bar{z}-\bar{s}z)=0 \nonumber \\
C1:&& g_v-2\frac{\alpha_v}{\alpha}g+r\bar{z}z=0 \nonumber \\
C2:&& f_u-2\frac{\alpha_u}{\alpha}f+r\bar{w}w+iea(s\bar{w}-\bar{s}w)
+e^2a^2\bar{s}s \nonumber \\
\nonumber \\
M1:&& a_v-\frac{\alpha^2 q}{2r^2}=0 \nonumber \\
M2:&& q_v-\frac{1}{2}ier^2(s\bar{z}-\bar{s}z)=0 \nonumber \\
\nonumber \\
S:&& rs_{uv}+fz+gw+iearz+ieags+ie\frac{\alpha^2 q}{4r}s=0.
\end{eqnarray}

\subsection{The initial conditions}
Having specified the equations to be solved we turn to discuss the
formulation of the initial conditions. The physical situation we
wish to describe is the gravitational collapse of a shell of
in-falling charged matter.We choose the domain of integration to
be a rectangle in the u-v plane (see section 3.2, where we
describe the numerical scheme). We can expect the \EH to be inside
the domain of integration, but not the \CH since it is located at
infinite null coordinate v (We are not using Kruskal-like
coordinates). we can However approach the  \CH asymptotically at
large values of v.

\begin{figure}[h]
\begin{center}
\includegraphics[width=8cm,height=9cm]{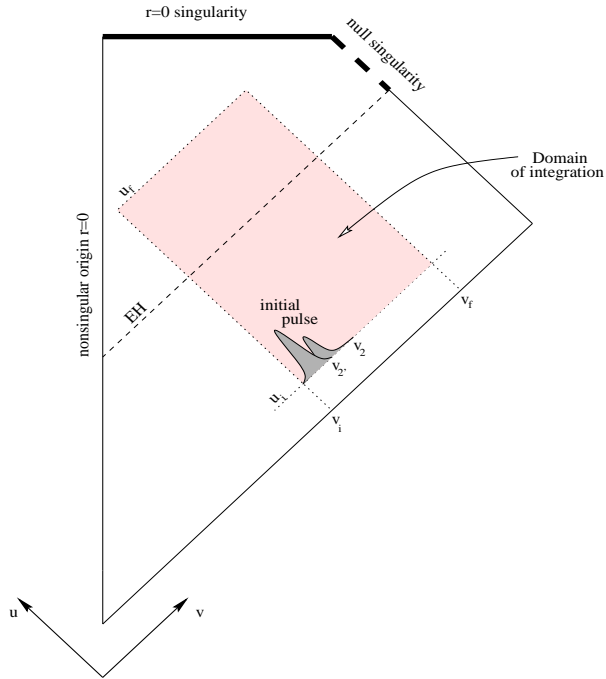}
\end{center}
\caption{Illustration of the domain of integration in relation to the expected
structure of the space-time. The \CH cannot be actually included in the domain
because it is in null infinity. Taken from \cite{evgenyfig}  \label{FIG domain}}
\end{figure}

When we come to specify the metric functions on the initial
hypersurface, we must fix the coordinate gauge freedom mentioned
in section 2.1. In a Minkowsky space-time we usually choose
$u=t-r, v=t+r$. Space-time around a gravitating spherical shell is
flat in two regions - inside the shell and at asymptotically large
radii. Since we have one constraint equation, C1, that relates
$r$,$\alpha$ and the scalar field on the initial hypersurface, we
can choose two of these functions arbitrarily. We would like to
take $r$ to be linear with $v$ in order to reflect the fact that
we start our integration far from the \EH, where the metric is
nearly flat. The fact that this is only an approximation will be
pronounced by the deviation of $\alpha$ from its flat-space value.
We can further specify the shape of the matter shell on the
initial hypersurface, and integrate C1 to obtain $\alpha$ on that
surface. Using the conventional flat-space dependence of $r$ on
$u$ and $v$,
\begin{equation}
r_i=\frac{v_i-u_i}{2},
\end{equation}
we conclude that $g_i=0.5$, and given an initial distribution of the scalar
field that we will soon elaborate on, we get from C1 the following ODE for
$\alpha$:
\begin{equation}\label{EQ initial constraint}
\frac{\alpha_v}{\alpha}=r\bar{z}z.
\end{equation}
We choose for the initial field distribution a compactly supported function,
which is basically one half period of a cosine. Studying equation M2 we
see that in order to have a non-vanishing charge distribution we must use for
the real and imaginary part of the field two pulses offset in v by some amount.
The exact form we choose is:
\begin{eqnarray}\label{EQ initial field}
Re[s_i] &=& \left\{ \begin{array}{cc}
              0 & v_1>v>v_2 \cr
              Ar(\frac{1+cos(\pi\frac{v-v_c}{\Delta v})}{2})^2 & v_1<v<v_2
             \end{array} \right. \nonumber \\ \nonumber \\
Re[s_r] &=& \left\{ \begin{array}{cc}
              0 & v_1'>v>v_2' \cr
              Ai(\frac{1+cos(\pi\frac{v-v'_c}{\Delta v'})}{2})^2 & v_1'<v<v_2'.
             \end{array} \right.
\end{eqnarray}
Having specified the two fields $r$ and $s$ on the initial
hypersurface, we can analytically derive $z$ and $g$ (which are
simply $g=\half$ and $z=\frac{\partial s}{\partial v}$), and
numerically integrate the remaining quantities $f,\alpha,a,q$ and
$w$ using equations $E1,C1,M1,M2$ and $S$ respectively.

We also need to specify the boundary conditions on the ray
$v=v_i$. Since the metric on the boundary, which is always inside
the collapsing shell, is flat, we can select the conventional
flat-space values for the fields there:
\begin{eqnarray}
r &=& \frac{v_i-u_i}{2} \nonumber \\
\alpha &=& 1 \nonumber \\
s &=&z=q=A= 0 \nonumber \\
\end{eqnarray}
In selecting these values for the metric functions on the boundary we specify
the physical meaning of the metric functions: The choice of $\alpha$ implies
that $\frac{u+v}{2}$ is the proper time for an observer located at the origin,
while $r$ is the proper surface area of a sphere of a radius $r(u,v)$.

For monitoring the mass content in our space-time we use the mass function,
\begin{equation}
m(u,v)=\frac{r}{2}(1+\frac{q^2}{r^2}+4\frac{r_ur_v}{\alpha^2}),
\end{equation}
which represents the total mass in a sphere of radius $r(u,v)$.

\section{The numerical scheme}
\subsection{Domain of integration}
The main consideration in choosing the domain of integration is
whether or not to include the origin. The origin will be important
when discussing critical phenomena \cite{choptuik}, however we are
only interested in effects that occur at large values of v and at
a finite radius, so we can choose the domain to be a rectangle as
shown in \FIG{domain}. The advantage of this approach is that the
origin, while being regular, is a coordinate singularity, and this
would force us to make a series expansion of the fields near the
origin in order to maintain numerical precision.
\subsection{Numerical Algorithm}
We construct a 2-dimensional grid in the v,u space and
integrate in increasing v and u direction. Since our
equal-coordinate surfaces lie on the light-cone, there are no
Courant-like stability limits on the step size. We implement a
Runge-Kutta like scheme by making a trial step to the middle of
the cell and using this information to make a $2^{nd}$ order full
step. We have found this scheme to be simple and efficient.

We use equation E1 to evolve r,f and g. E2 is used to evolve
$\alpha$, and S evolves s,z and w. Finally M1 and M2 evolve a and
q respectively. Note, that this scheme is a Free Evolution scheme,
since only the dynamical equations are used for evolving the
space-time. The constraint equations are used for
monitoring the accuracy.

\subsection{Adaptive grid refinement}
A major problem in black hole numerics is the behavior near the
\EH. Although the double-null coordinates we have chosen ensure
regularity of the fields even as we cross the horizon, there is
nevertheless a fundamental difficulty in following the evolution
numerically because of the physical behavior near the Horizon.
Consider two outgoing null rays starting at the origin,
slightly above and below the Event Horizon. One is destined to
escape to infinity while the other will remain trapped inside the
Horizon. This means that regardless of how close the two rays were
initialii, their distance will diverge as
their advanced time v grows. The numerical implication is that the
metric function $f=r_u$ diverges along the Event Horizon, and $r$
becomes ``discontinuous'' asymptotically. If we want to
maintain a fixed relative change in the area coordinate r, we have
to make the grid denser as we approach the horizon. The step size
has to decrease with the distance form the horizon. This decay
turns out to be exponential, and the exponent increases as we
increase $v_{max}$.
 It would be best to change both $\Delta u$ and $\Delta
v$ as we approach the horizon and reach large values of v, respectively, but we
chose in this work to change only $\Delta u$ because while this is
significantly simpler to implement, it gives reasonable results.
\begin{figure}[t!]
\begin{center}
\includegraphics[width=10cm,height=6cm]{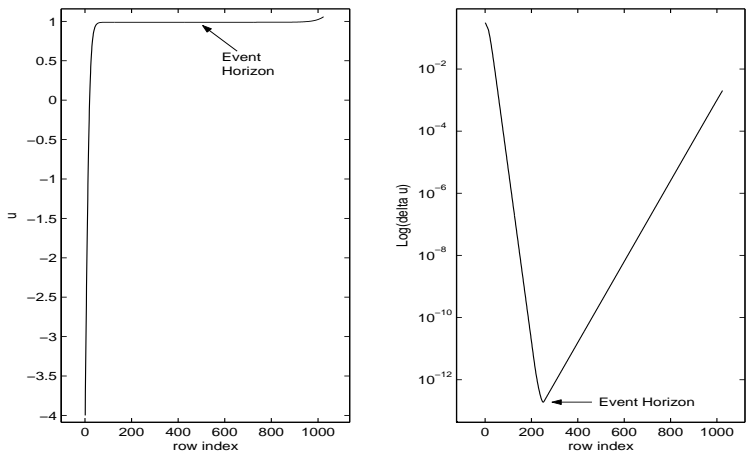}
\caption{Adaptive grid refinement. The left figure features the
  development of u along a typical grid, showing how most of the
  computational effort is concentrated at the neighborhood of the
  Horizon. The figure on the left is the difference $\Delta u$,
  showing that $\Delta u$ changes by many orders of
  magnitude. \label{FIG du}}
\end{center}
\end{figure}
The choice of having the refinement algorithm maintain $\frac{\Delta r}{r}$
constant, instead of some other local indicator of accuracy, is not trivial and
was made after trial and error indicated that it was the best strategy, and
also because r is always finite in the region of interest so it provides a good
scale for measuring error. This scheme dictates, for relatively high values of
$v_{max}\approx 100M$, a condensation of about 10 orders of magnitude in
$\Delta u$ in the vicinity of the \EH relative to the initial value of $\Delta
u$. This is illustrative of the difficulty in numerically crossing the \EH at
large values of v. \FIG{du} shows an example of the evolution of $\Delta u$ on
a typical space-time.

\subsection{Numerical tests}
Our first test would have been a comparison of the
numerical results with a known analytical solution. Unfortunately,
there are no suitable analytical solutions. Therefore, we
will have to check our code by other means. We begin with a test
of convergence, i.e. we verify that the various fields converge to
some value in the expected order of convergence. We evolve the
same initial conditions on three grids, $g_1, g_2$ and $g_3$ which
have fixed grid densities (in both u and v directions) $d_1, 2d_1$
and $4d_1$. We need three grids because we do not know the exact
solution, so we can compare only the relative error. In
\FIG{converge} we plot, for example,  the functions
$Re[\psi_1-\psi_2]$ and $4Re[\psi_2-\psi_3]$. We expect, and
verify, that the two curves be nearly but not exactly the same,
because higher order terms also have a small contribution to the
error. The data in this figure is from a single row($u=const$) but
it is representative of the entire grid.
\begin{figure}[h]
\begin{center}
\includegraphics[width=10cm,height=6cm]{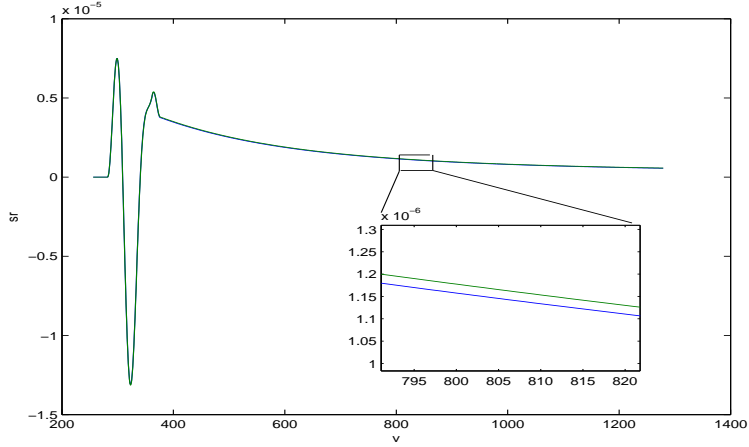}
\caption{$Re[\psi_1-\psi_2]$ \& $ 4Re[\psi_2-\psi_3]$ for the real
part of the scalar field on one of the outgoing rays. The two
curves are almost indistinguishable, which shows that the scheme
converges nicely to
  $2^nd$ order \label{FIG converge}}
\end{center}
\end{figure}

Although convergence is crucial, it still does not guarantee that the results
are correct, because if there is an error in the equations, either in the
original equations or as they are implemented in the code, the scheme will
still converge but to the wrong solution. We are in a unique situation where
the constraint equations provide us with a measure of consistency. The Einstein
equations preserve the constraints, in the sense that if the initial conditions
satisfy the constraint equations, then they will be satisfied also under
evolution of the dynamical equations. This is true analytically, but since our
solution is numerical we cannot solve exactly either the constraint equations
on the initial hypersurface or the dynamical equations through the evolution.
Therefore we can only demand that the \textit{error} in the constraint will
converge to zero as we make our solution more accurate by condensing the grid,
i.e. $C \stackrel{\tiny{stepsize\rightarrow 0}}{\verylongrightarrow} 0$ . This
is verified in \FIG{constraint}.
\begin{figure}[h]
\begin{center}
\includegraphics[width=10cm,height=5cm]{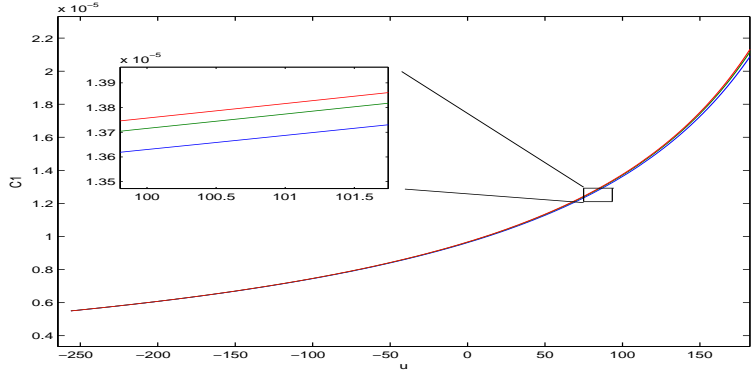}
\end{center}
\caption{The maximum constraint error along outgoing rays
$max_v(C_1)$, $max_v(4 C_2)$ and $max_v(16 C_3)$ as a function of
u on the same 3 grids as in \FIG{converge}. The three curves are
again almost indistinguishable, showing that the constraint
violation displays $2^{nd}$ order convergence to zero. The initial
error in the constraint grows with u because of unavoidable
accumulation of numerical errors. \label{FIG constraint}}
\end{figure}

Next we check that our code reproduces known features, such as the location of
the horizons, mass and charge conservation etc. Comparing the location of the
horizons to the \RN solution's $r_\pm=M\pm \sqrt{M^2-q^2}$ we find a
correspondence of about one part in $10^5$ between the computed and expected
radii.

\begin{figure}[t!]
\begin{center}
\includegraphics[width=10cm,height=10.8cm]{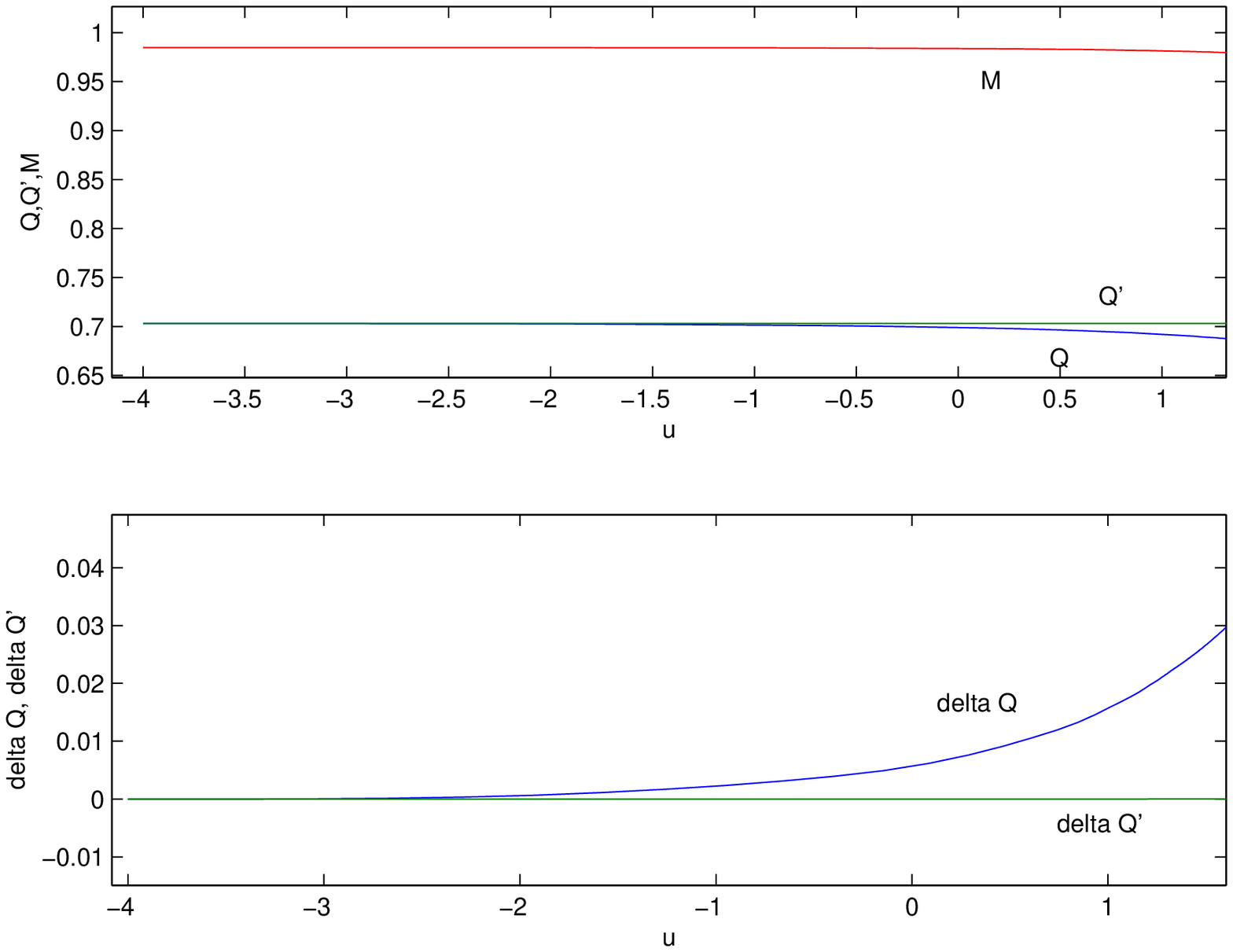}
\end{center}
\caption{Global charge and mass conservation. Upper
  figure: Q is the charge $q(u,v)$, Q' is the charge, compensated for the outgoing current density at the last grid
  point (outgoing flux). Lower figure: the relative change
  $1-\frac{Q}{Q_0}$ and $1-\frac{Q'}{Q_0}$, which is four
  orders of magnitude smaller.  \label{FIG conservation} }
\end{figure}

Finally we observe that global mass and charge are conserved. Actually this is
not exactly the case since some of the field is scattered by the gravitational
and electromagnetic potential as the shell approaches its gravitational radius,
but this is in most cases (when $Q<M$) an insignificant amount and the
conservation laws seem to hold. \FIG{conservation} shows the mass M and the
charge Q at the last grid-point vs. u. Also shown is Q' which is Q compensated
for the charge lost at the last grid point due to outgoing flux:
\begin{equation}
Q'(u)=Q(u)+\int_{u_0}^u{r^2(u',v_{max}) Ju(u',v_{max})du'},
\end{equation}
which follows from the conservation equation for the electromagnetic
current:
\begin{equation}
J^\mu_{\ \mu}=0.
\end{equation}

It should be noted that the mass inside the domain of integration
is not generally conserved, since the reflected waves carry off
some mass, as well as charge. The subject of mass conservation is
more delicate than charge conservation since the total mass
contains also the Energy-Momentum of the gravitational field,
which doesn't enter into the Noether current related to the
Energy-Momentum tensor. We will refrain from dealing with this
topic here.
\subsection{Numerical error}
To summarize, we can obtain a measure of the relative numerical errors from the
tests we discussed. The first test is to use the convergence test and compare
the difference in the fields between different grids. There is a difficulty
here because this test can only be performed without using adaptive grids. This
substantially limits the precision that can be attained in this test, but as
\FIG{delta r} illustrates, even with this limitation we get good results: $\sim
10^{-3}$ maximum relative error.As we indicate in the next paragraph,  the
error will be lower by an order of magnitude or more if we employ the grid
refinement algorithm.

\begin{figure}[t!]
\begin{center}
\includegraphics[width=10cm,height=8cm]{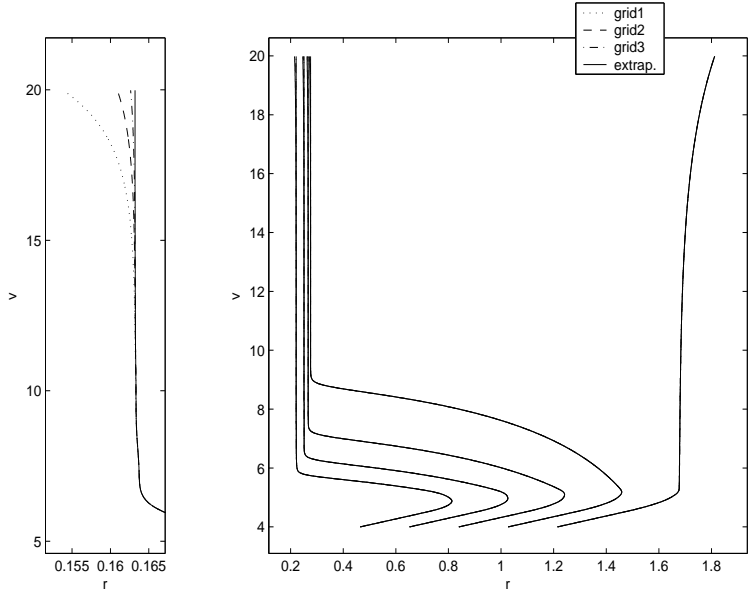}
\end{center}
\caption{Right panel: The metric function r vs. v for several outgoing null
rays, on 3 different grids $g_{1,2,3}$ with grid spacing 4h, 2h and h
respectively. The three grids are compared to a solution extrapolated from
them, taking into account the $2^{nd}$ order convergence. This solution is much
more accurate than the three others and can be taken to as the ``accurate''
solution for the purpose of comparison. The 4 curves are almost
indistinguishable. Left panel: detail of the leftmost ray \label{FIG delta r}}
\end{figure}

The next test is charge conservation. We can see in \FIG{Qtag} that the charge
conservation error is less than $10^{-4}$ for the most dense grid. We can also
see that the error converges to zero. These results are taken from refined
grids, because in this case we are not comparing rows against each other so we
don't need to match rows exactly in u value. This test was done with
$v_{max}=20M$. In this case the same calculation on an unrefined grid gave an
error larger by an order of magnitude. This ratio will be larger as we increase
$v_{max}$, making the use of grid refinement more and more crucial.

\begin{figure}[h!]
\begin{center}
\includegraphics[width=9cm,height=6cm]{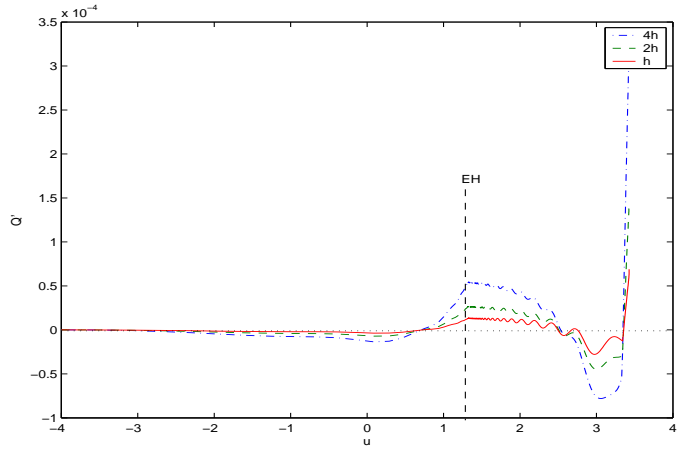}
\end{center}
\caption{$1-\frac{Q'}{Q_0}$ for three different grids vs. u. The
dashed line indicates the position of the EH. The legend indicates
the relative step size used in each grid. \label{FIG Qtag}}
\end{figure}

The last test we perform is constraint violation. We can split C1 in
\EQ{notations} into two terms and compare their relative error:
\begin{eqnarray}
A&=&g_v-2\frac{\alpha_v}{\alpha}g \\
B&=&-r\bar{z}z.
\end{eqnarray}
Now we can take $\frac{|A|-|B|}{|A|+|B|}$ to be the relative error. However
there is a problem since $|A|+|B|$ has no definite scale and it often
intersects zero, so the error behaves very badly. Therefore we will compare
$(A-B)$ in each $u=const$ segment to the r.m.s. of $(A+B)$ along the same
segment. The results are given in \FIG{constraint error}, and show a maximum
error level of $\sim 10^{-5}$ in the finest grid used.

\begin{figure}[t!]
\begin{center}
\includegraphics[width=9cm,height=6cm]{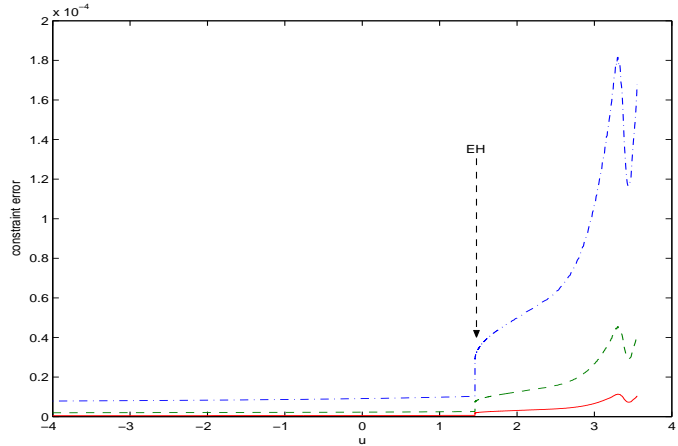}
\end{center}
\caption{The relative error in the constraint,
$max_v(\frac{A-B}{\sqrt{<(A+B)^2>_v}})$ vs. u. The three
curves are for the same 3 grids as in \FIG{delta r}.  \label{FIG constraint error}}
\end{figure}

\section{Results}
\subsection{Formation of the black hole}
\begin{figure}[h!]
\begin{center}
\includegraphics[width=10cm,height=7cm]{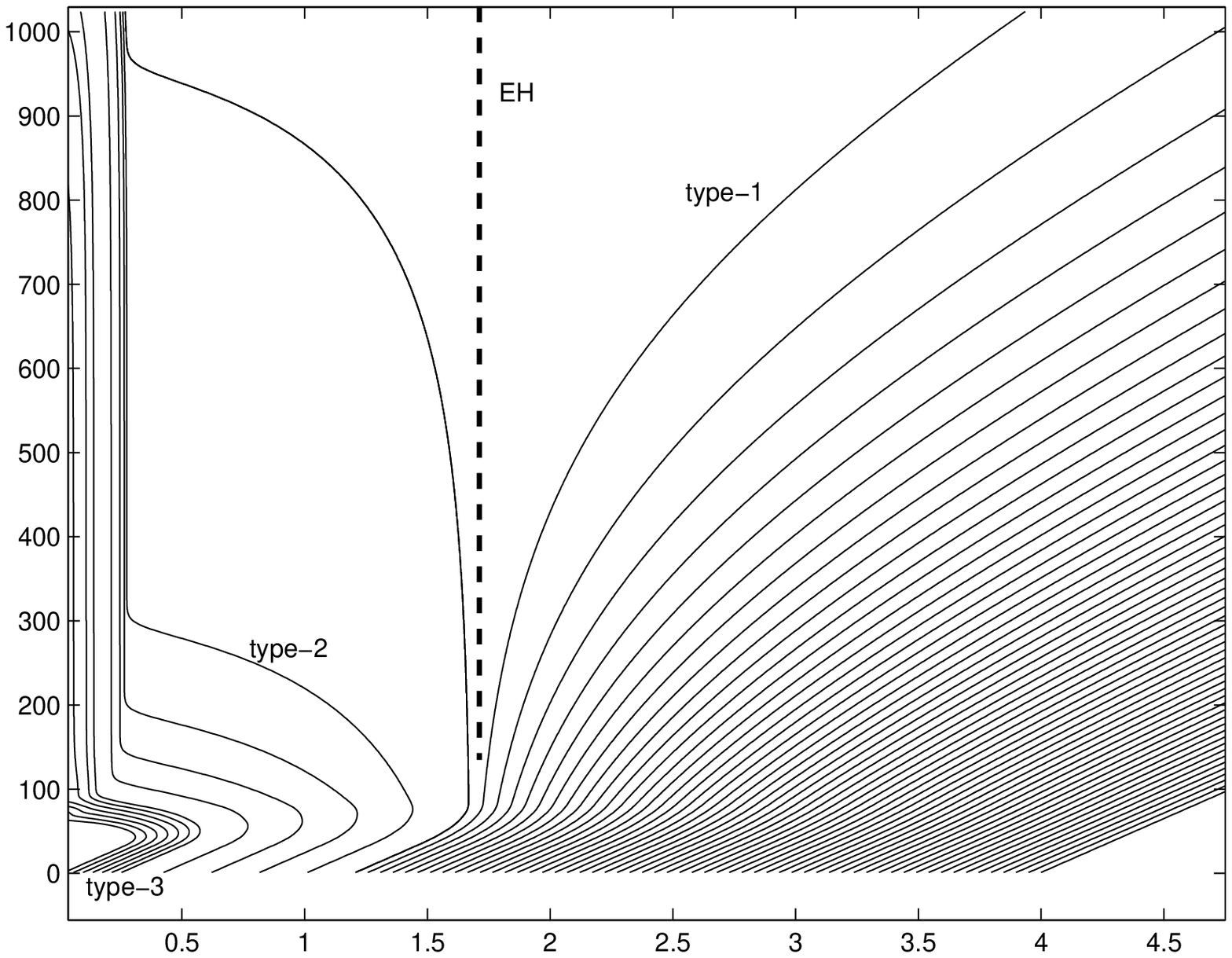}
\includegraphics[width=10cm,height=5cm]{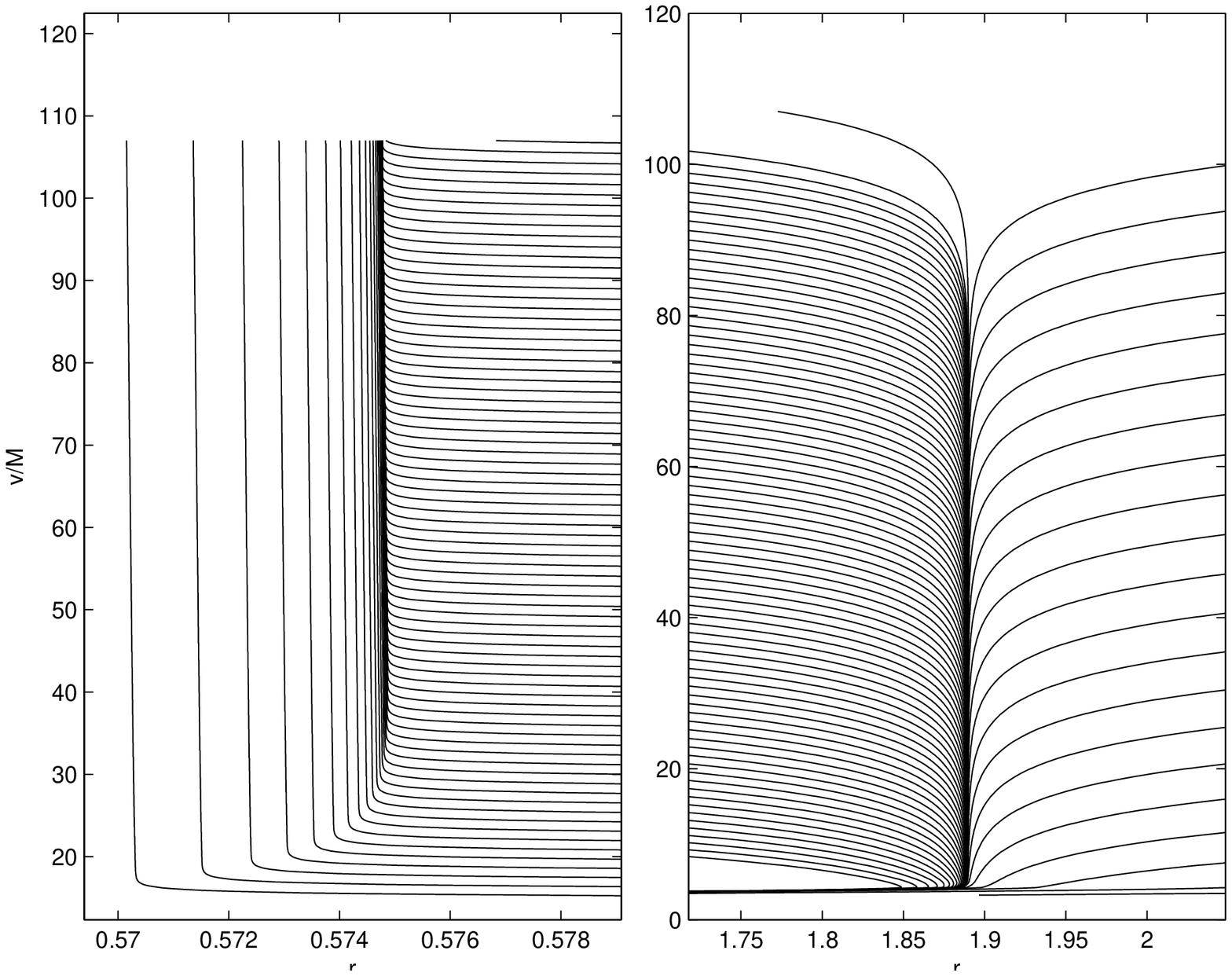}
\end{center}
\caption{Null rays. The top figure shows the
  entire domain of integration, and the three classes of rays: type-1
  rays reach null-infinity, type-2 approach the \CH and type-3 terminate at
  the singular origin. The dashed line signifies the approximate
  location of the Apparent Horizon. the two lower figures focus on the
  vicinities of rays approaching the \CH (left) and the \EH (right). \label{FIG rv}}
\end{figure}
We begin \FIG{rv} with a null-ray diagram giving an overview of
the space time. The diagram shows the formation of an \EH and then
a Cauchy Horizon at asymptotically large v. Looking
closely at the null rays approaching the Cauchy Horizon we see
that it is not stable, as the asymptotic ($v\rightarrow \infty$)
value of $r$ decreases as $u$ increases. Finally at large enough
values of u the rays reach the origin ($r=0$), which signals the
appearance of a space-like singularity.

\subsection{charge Dynamics}
\begin{figure}[b!]
\begin{center}
\includegraphics[width=12cm,height=6cm]{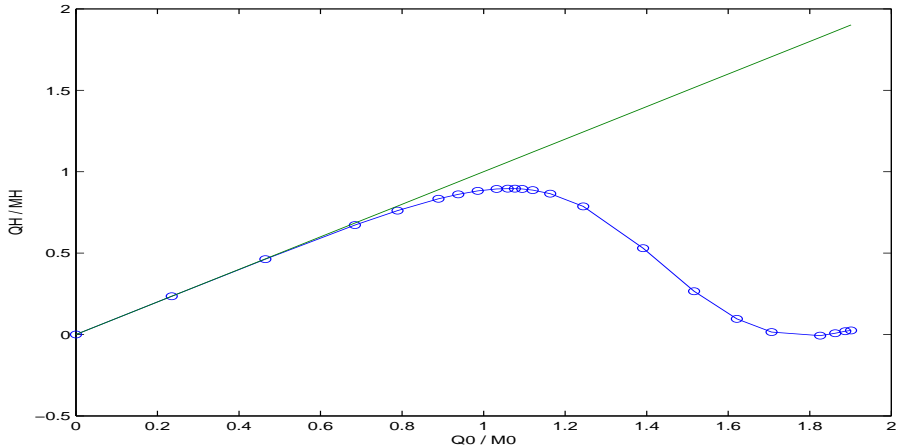}
\caption{The charge to mass ration on the \EH for different initial
  conditions. \label{FIG qm}}
\end{center}
\end{figure}

The Cosmic Censorship conjecture \cite{censor} states that all
singularities in nature are contained within an \EH, so there are
no ``Naked singularities''. It follows from this that it is
impossible to overcharge a black hole (i.e. to increase its
charge-to-mass ratio above 1) since there is no black hole
solution with this ratio. It is interesting to check if this holds
in our collapse scenario by increasing the initial charge of the
collapsing shell. It can be seen in \FIG{qm} that, for a certain
initial configuration, as we increase the coupling constant e, and
with it the initial charge to mass ratio, the final black hole
charge to mass ratio reaches a maximal value smaller than 1 and
then decreases. The physical mechanism that causes this behavior
is the electrostatic repulsion of the outer parts of the shell,
which increases relative to the gravitational pull they experience
as the charge increases, and causes them to be reflected back
towards large radii at the early stages of the collapse and
radiate part of the initial charge away. This can be understood as
a scattering process, where at each instant in time the shell
encounters some potential barrier with certain reflection and
transmission coefficients. As long as the charge-to-mass ratio
will be greater than unity, no horizon will form, and charged
matter will be able to escape. Ordinarily the charge and mass
become equal some finite time before the shell crosses it's
gravitational radius and a Horizon forms, so the final
charge-to-mass ratio decreases further and ends up smaller than
unity.

\begin{figure}[t!]
\begin{center}
\includegraphics[width=12cm,height=6.5cm]{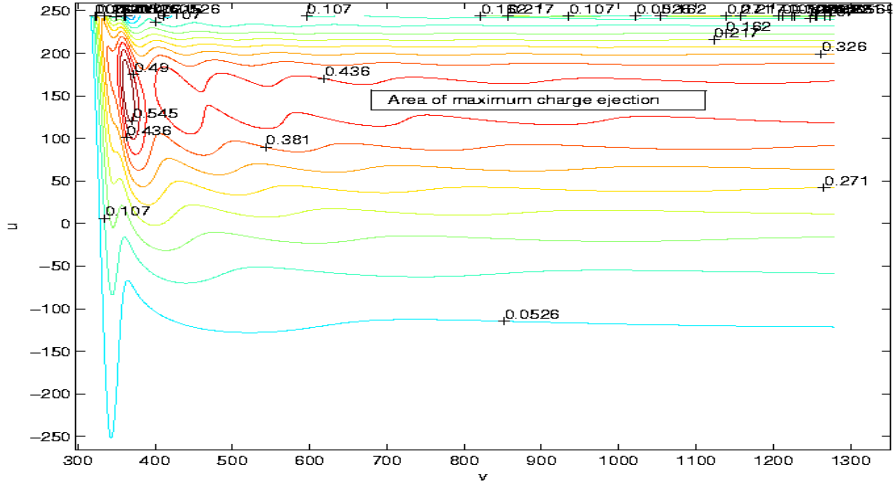}
\caption{The dynamical ejection of excess charge. Most of the
charge is removed well before the shell reaches the horizon (upper
edge, $u\approx 250$), at $u\approx 150$.\label{FIG J contour}}
\end{center}
\end{figure}

This process can be seen by inspecting the movement of charges on
the computational grid. The outgoing current density, in our
coordinates, is proportional to the u component of the 4-current
$J_\mu$. Note that this is a matter of interpretation since
normally the charge density would be the time-like component of
the 4-current, and the current density would be the space-like
component. In our coordinate system u and v are both null, but we
treat v and u as the ``spatial'' and ``temporal'' coordinates
because we are dealing with an in-falling shell. \FIG{J contour}
shows a contour of this quantity for the case
$\frac{Q_0}{M_0}=1.7$ .

\subsection{Wave tails}
Next we observe the mechanism behind the No-Hair theorem, mentioned in
chapter 1. We inquire how the black hole radiates away the ``hair'', i.e. any
feature of the collapsing matter except it's charge, mass and angular momentum.
We want to compare this decay to theoretical results predicting a certain power
law decay of the tails at late times. Note that we perform this analysis along
the Event Horizon, which is a null hypersurface with a constant radius.
Hod\&Piran \cite{HPtail3} predicted in this case the following behavior on the
Horizon:
\begin{equation}
\psi\sim e^{i\frac{eQ}{r_+}} \ v^{-\beta} \ \ ;\ \
\beta=1+\sqrt{(2l+1)^2-4(eQ)^2}, \label{EQ tail index}
\end{equation}
where $l$ is the multipole moment of the scalar perturbation. In our case $l=0$
because we deal with spherical symmetry. The first term in \EQ{tail index} is
an oscillation term with angular frequency $\frac{eQ}{r}$. The second term is a
power-law decay of the field magnitude. Typically the field magnitude on the
horizon begins to decay with Quasi-Normal ringing (exponentially decaying
oscillations) which decays exponentially, Then the power-law tail sets in and
continues to asymptotically large advanced time v where it dominates the
behavior of the field. This can be clearly seen in \FIG{tail}.
\begin{figure}[h!]
\begin{center}
\includegraphics[width=14cm,height=14cm]{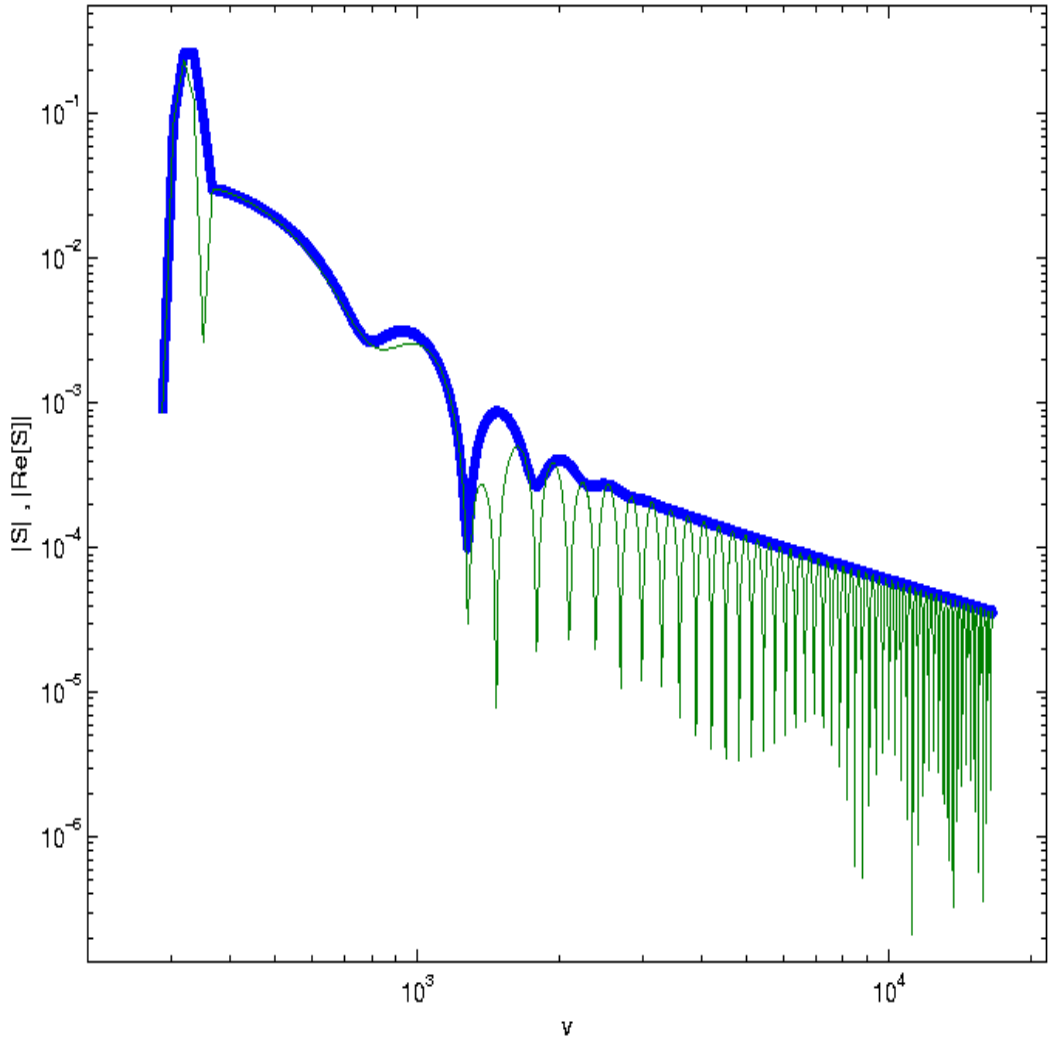}
\caption{A typical configuration of the scalar field amplitude on the
  horizon, showing Quasi-Normal ringing followed by a power-law decay
  (the ``tail''). The thick line is the amplitude of the field, the
  thin line is the absolute value of the real part of the field. \label{FIG tail}}
\end{center}
\end{figure}

The above expression for the oscillation and power law index were checked over
a range of the parameter eQ, by changing e and leaving all other parameters
constant. The results in \FIG{tail beta} show good correspondence with both
terms in \EQ{tail index}.
\begin{figure}[h!]
\includegraphics[width=12cm,height=6cm]{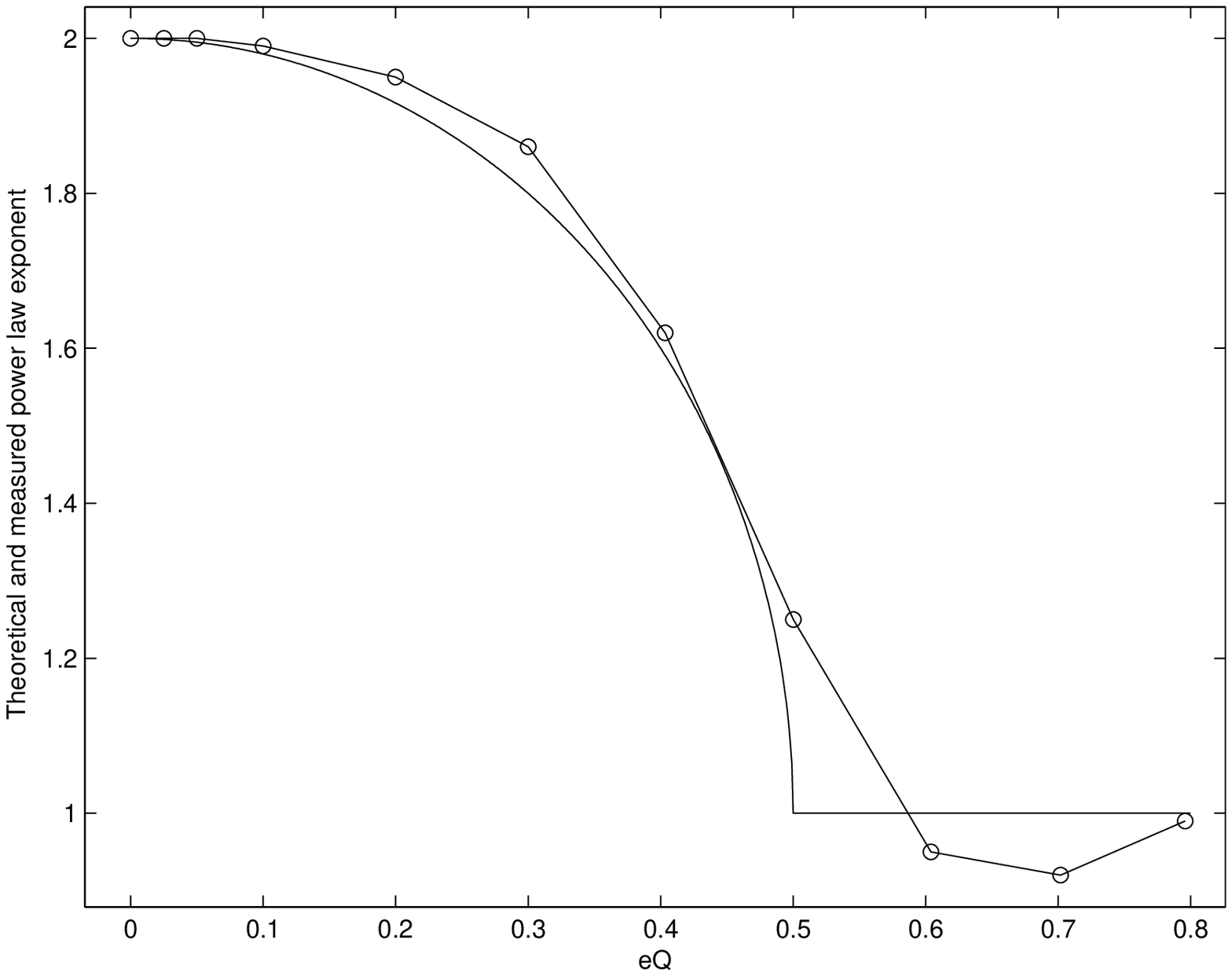}
\includegraphics[width=12cm,height=6cm]{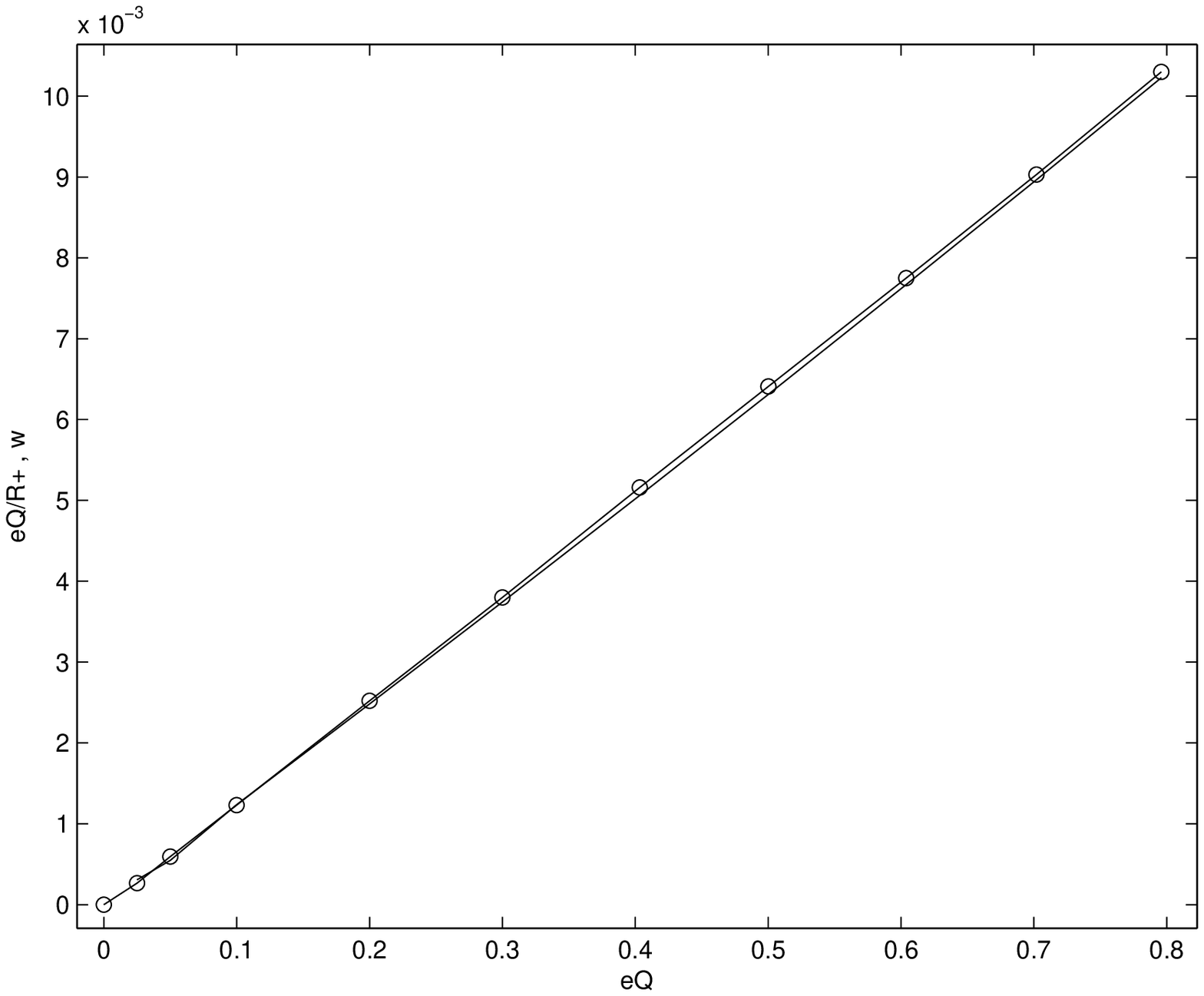}
\caption{Upper figure: The real part of the power law index for
  spherical charged perturbations versus the theoretical
  prediction. Lower figure: The phase oscillation frequency versus the
  theoretical value, $\frac{eQ}{r^+}$. \label{FIG tail beta}}
\end{figure}

\subsection{The \CH}\label{SEC CH}
\begin{figure}[p]
\begin{center}
\includegraphics[width=11cm,height=7cm]{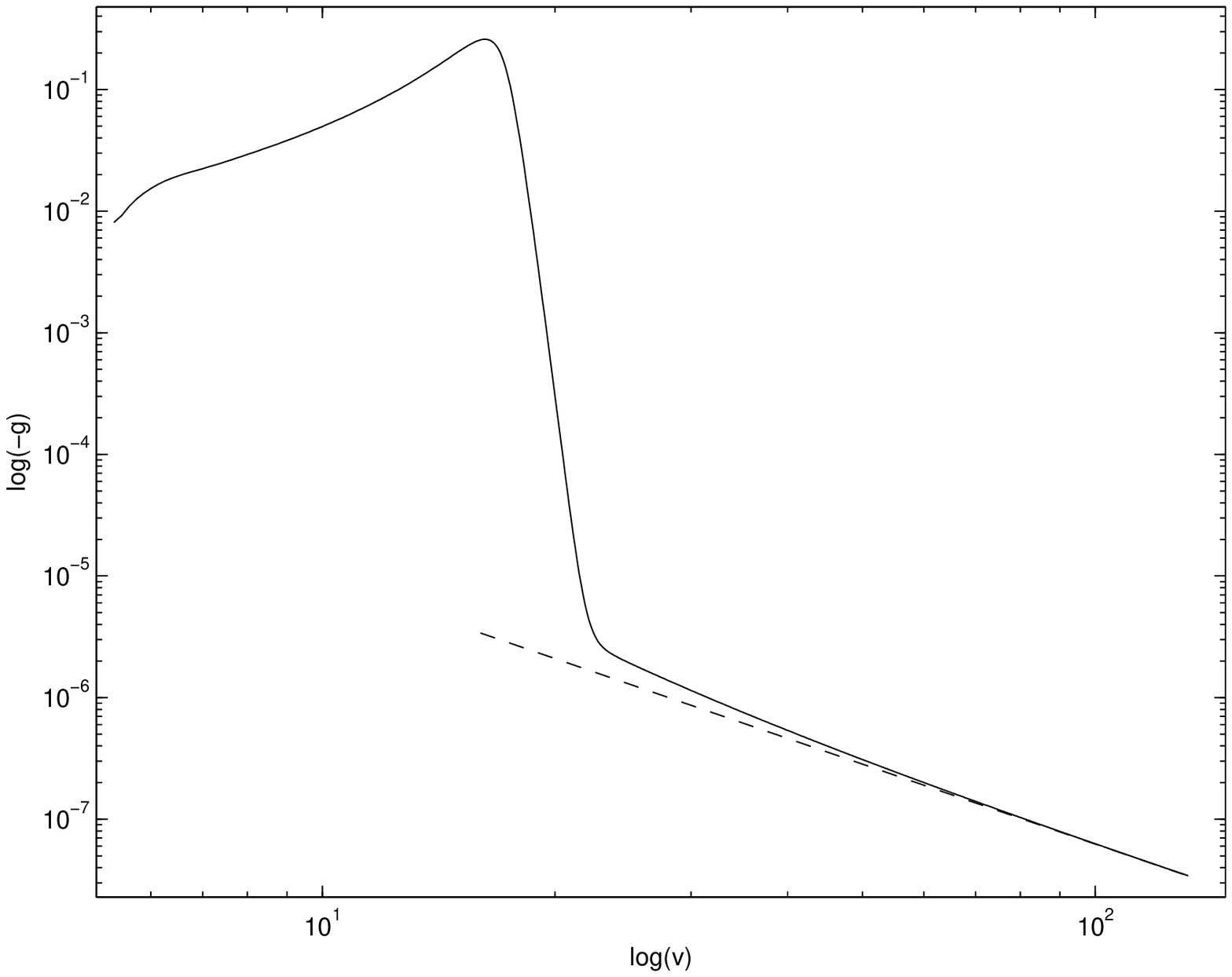}
\caption{The behaviour of $r_{v}$ along a null ray approaching the
  \CH. The power-law decay can be seen clearly at late times.
\label{FIG CH powerlaw1}} \vspace{8mm}
\includegraphics[width=11cm,height=7cm]{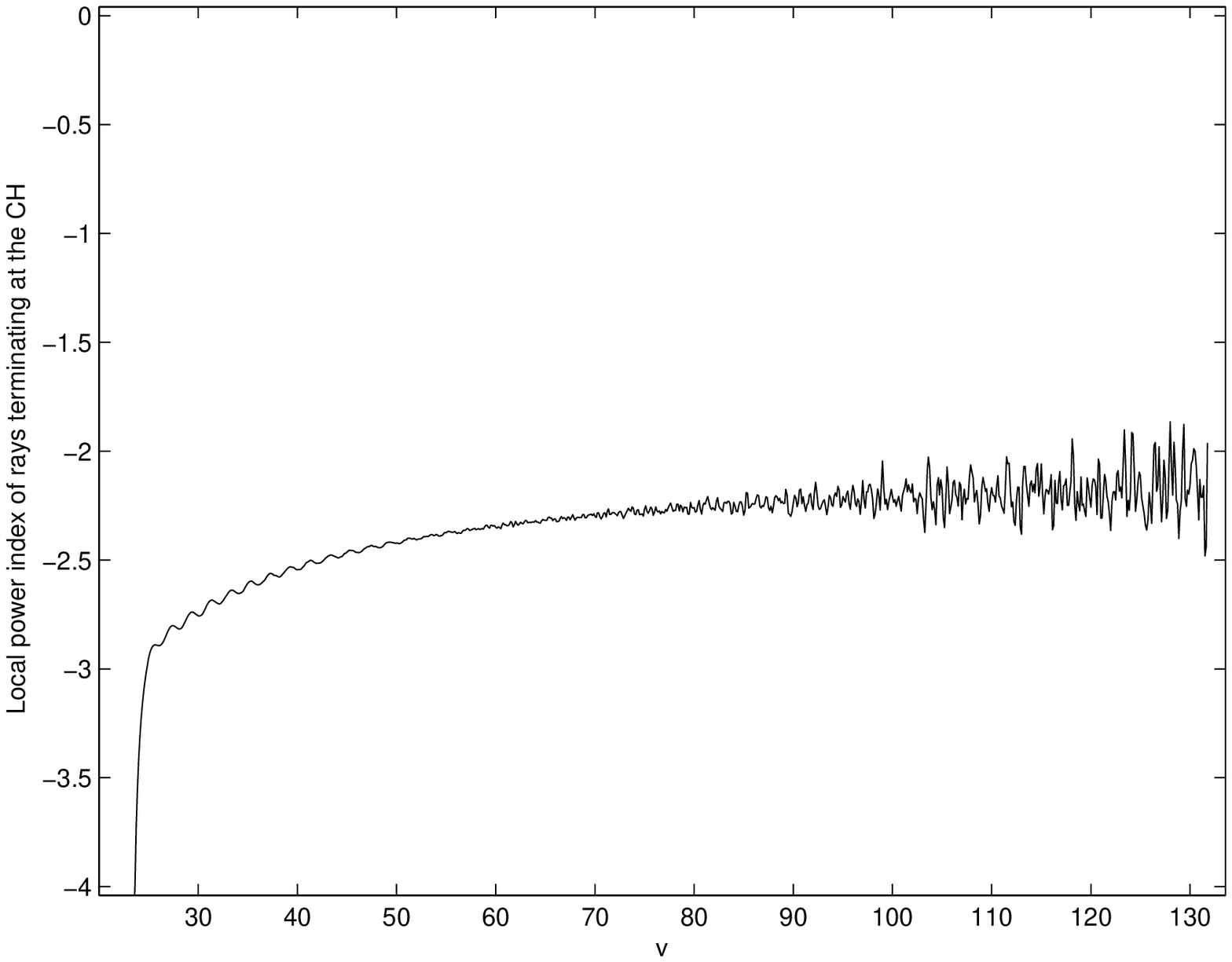}
\caption{The local power-law index of $r_v$ at the approach to the \CH. $r_v$
behaves like $v^{-2}$ at large v, signifying that r itself approaches a
constant value as $v^{-1}$. The noise at large v is only a numerical artifact.
\label{FIG CH powerlaw}}
\end{center}
\end{figure}

We now turn to examine the behavior of the space-time near the Cauchy Horizon.
First we try to verify that null rays of type-2 actually reach an
asymptotically fixed radius, rather than assuming a course that would
ultimately bring them, at very large values of v, to the origin. In order to do
this we must extend our domain of integration, so that we can observe the decay
of $r_v$ over at least one order of magnitude and establish the asymptotic
behavior. In \FIG{CH powerlaw} and \FIG{CH powerlaw} such an analysis is made,
showing that $r_v\sim v^{-2}$. This confirms the existence of a Cauchy Horizon.

The next thing we want to know is wether the \CH is singular, and if so how
much. One indication of a singularity is an exponential divergence of the mass
function. This can be seen in \FIG{inflation}, for a typical space-time. This
divergence is dominated by an exponential decay of $\alpha$ with v, which makes
the mass function diverge exponentially, because it depends on $\alpha$ as
$\alpha^{-2}$. This can be interpreted physically as an infinite blue-shift of
in-falling radiation at the Cauchy Horizon, since $\alpha$ is also $g_{tt}$ if
we shift to a time-space (1+3) coordinate frame. The theoretical prediction for
mass-inflation is $m\sim e^{\kappa v}$, where $\kappa$ is the surface gravity
on the inner horizon. For a static \RN black hole this is given by \cite{Brady}
\EQ{kappa}, and it agrees numerically to about $10-20\%$ with the measured
exponent.
\begin{equation}\label{EQ kappa}
\kappa=\frac{\sqrt{1-(\frac{Q}{M})^2}}{(1-\sqrt{1-(\frac{Q}{M})^2})^2}.
\end{equation}

\begin{figure}[t]
\includegraphics[width=12cm,height=10cm]{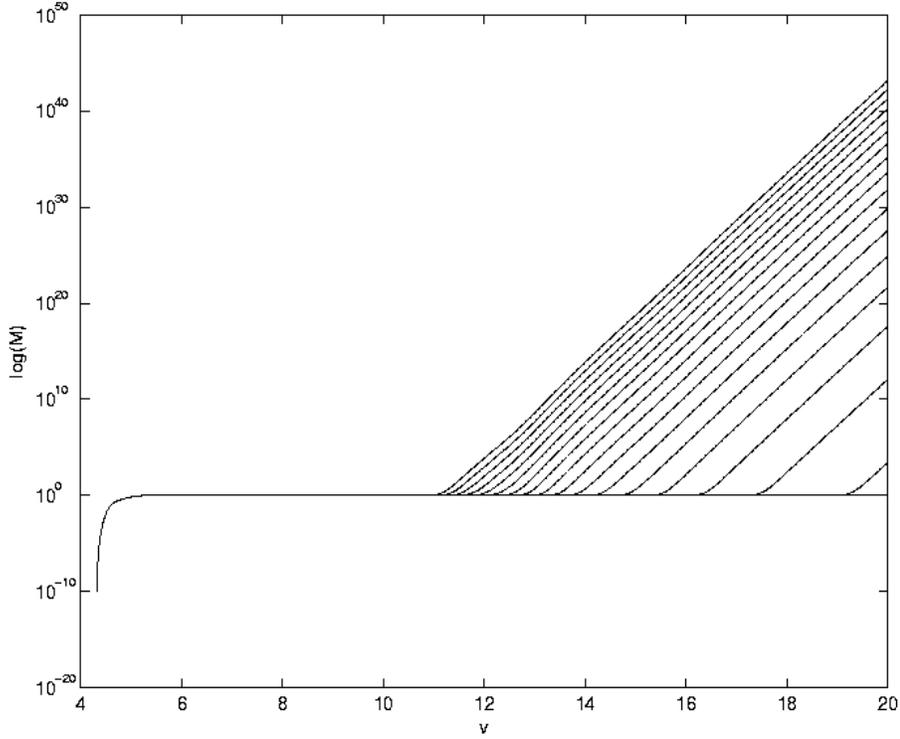}
\caption{Mass inflation near the CH. Each curve is for a constant u, u is
increasing from lower to higher lines. \label{FIG inflation}}
\end{figure}

The singularity also manifests itself in the Ricci curvature scalar, R. The
Ricci scalar is a purely geometrical entity, which can be written only with the
metric functions and their derivatives. However, we can use our dynamical
equations to express this quantity through the physical fields in the problem.
We arrive at this form:
\begin{equation}
R=-\frac{4}{\alpha^2}[(\bar w z+w\bar z)+iea(s \bar z-\bar s z)].
\end{equation}
R can be seen to diverge exponentially on the horizon while oscillating at some
definite frequency (see \FIG{ricci}). This behavior indicates the existence of
some kind of curvature singularity, but does not tell us if it is strong or
weak, i.e. if the tidal forces felt by an observer crossing it are infinite or
finite.
\begin{figure}[h!]
\includegraphics[width=12cm,height=7cm]{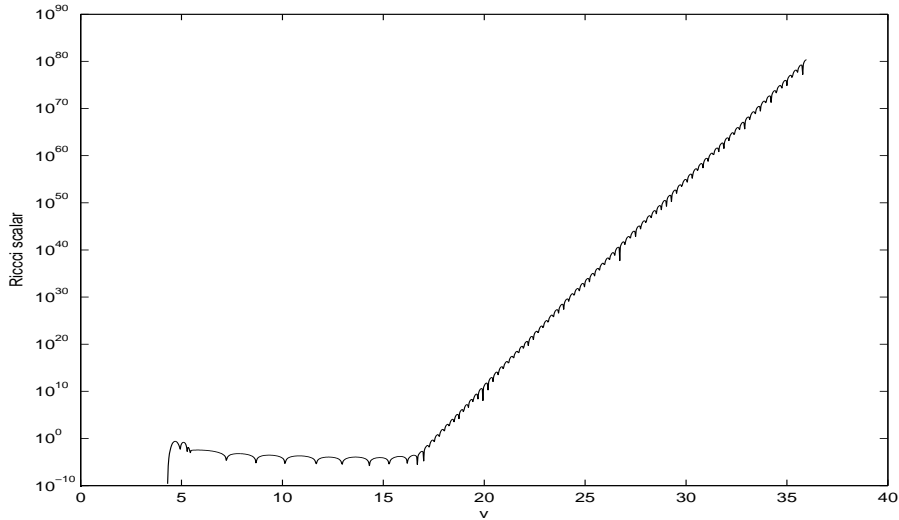}
\caption{The Ricci scalar on an outgoing null ray intersecting the Cauchy
Horizon. The cusps indicate points where R changes sign. \label{FIG ricci}}
\end{figure}
The divergence in R is caused by the exponential decay of $\alpha$, and the
oscillation is caused by the oscillatory behavior of the scalar field. We can
observe the frequency of this oscillation by looking at $R\alpha^2$ which
contains the oscillating term without the exponential divergence. \FIG{ricci
fft} shows this oscillation and it's Fourier transform. A strong oscillation
with period 0.5M is accompanied by a weaker component with period 0.6M, and an
oscillation with ``beat'' is produced.

It is interesting to note here, that as the curvature radius at the Horizon of
a black hole is on the order of magnitude of it's gravitational radius, which
is $\sim10^4m$ for a solar mass black hole, and the planck scale is
$\sim10^{-35}m$, the Ricci curvature can grow by approximately 40 orders of
magnitude before it reaches Planckian scales. This means that our analysis
brings us well into the Plackian regime, where classical GR begins to fail.
\begin{figure}[h!]
\includegraphics[width=12cm,height=14cm]{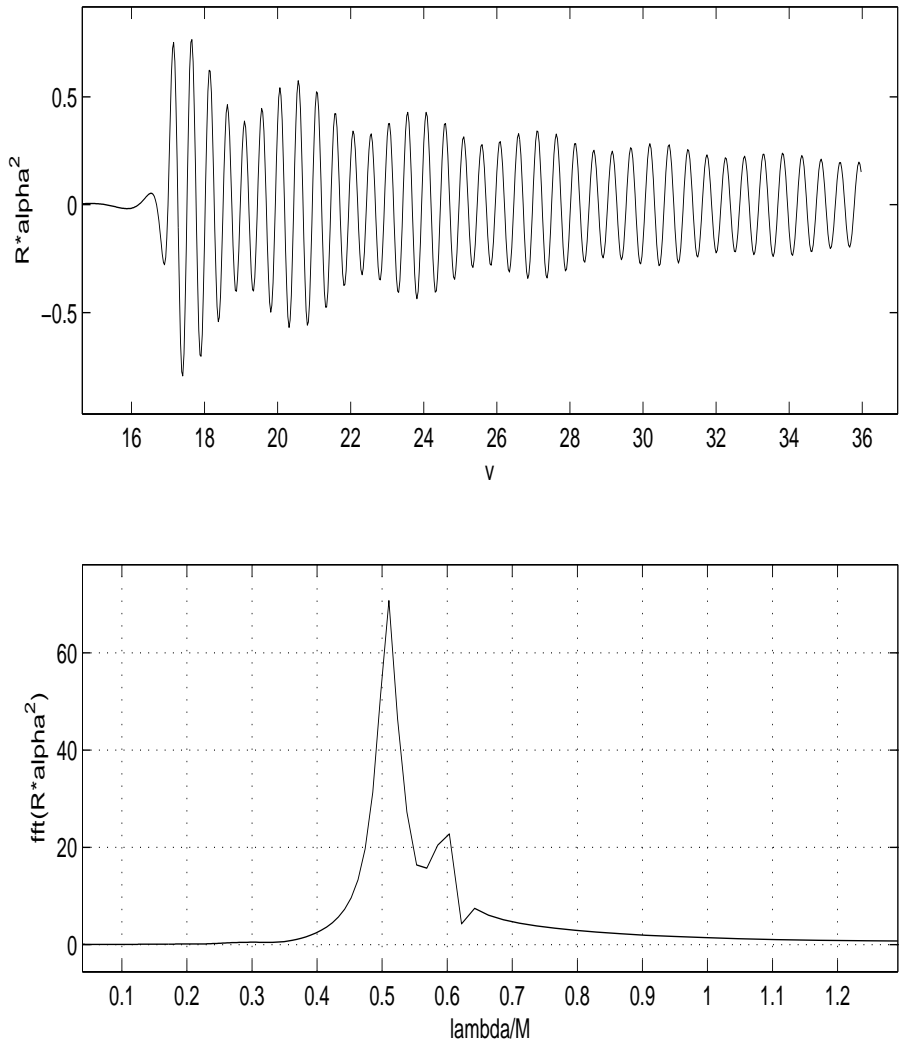}
\caption{The oscillatory part of the Ricci scalar. Top panel shows an
oscillation with a weak ``beat''. The bottom panel shows the Fourier transform
of the curve in the top panel. \label{FIG ricci fft}}
\end{figure}

\section{Summary and conclusions}
We have developed a numerical code that solves the coupled Einstein-Maxwell
equations in a dynamical collapse situation. This has enabled us to explore
phenomena involved in black hole formation that was previously handled by
analytic or numerical perturbative analysis. The difficulty in this kind of
numerical code is in maintaining small numerical errors in spite of extreme
gradients near the Event Horizon. We have solved this problem by using a
non-uniform grid that covers the difficult areas with dense grid points. The
scheme was shown to converge and give results accurate to better than than 1
part in $10^3$, even in areas deep inside the black hole.

We have shown that in a dynamical collapse of charged matter, some of
the charge is radiated away because of electrostatic repulsion and
scattering on the gravitational potential. This is in accord with the
Cosmic Censorship conjecture which forbids the charge in the Black
Hole to surpass it's mass. We also observed the radiative tails that
are left on the Horizon after the black hole is formed. These were
shown to have the well known structure of an initial decaying
oscillation (Quasi-Normal ringing) followed by a power law decay that
continues asymptotically.

Finally we ventured deep inside the black hole interior and examined the
properties of the inner Horizon. We have found that before being completely
destroyed and turning into a strong space-like (i.e. Schwarzschild-like)
singularity it behaves as a weak, null singularity. Since the \CH is located at
$v=\infty$ in our scheme we cannot reach it by numerical evolution. However
physically it is reached in a finite proper time by an in-falling observer
(This is true since $\alpha^2(u,v) dt$ which is the proper time differential
for an observer at $(u,v)$ decays exponentially as $v\rightarrow \infty$,
giving a finite lapse of proper time until reaching the Cauchy Horizon). The
weakness of the singularity thus leaves open the question of the traversability
of the ``Kerr Tunnel'', making it unclear wether it is physically possible for
matter to cross the CH into a another asymptotically flat region.

%%%%%%%%%%%%%%%%%%%%% BIBLIOGRAPHY %%%%%%%%%%%%%%%%%%%%%

\end{document}